\documentstyle[pra,aps,epsfig]{revtex}

\def\be{\begin{eqnarray}}
\def\ee{\end{eqnarray}}
\def\r{\rangle}
\def\l{\langle}

\begin{document}

\twocolumn
\draft

\wideabs{ 
\title{Quantum synthesis of arbitrary unitary operators} 
\author{B. Hladk\'y$^{1,2}$, G. Drobn\'y$^{1}$  and V. Bu\v{z}ek$^{1,3}$}
\address{$^{1}$ Institute of Physics, Slovak Academy of Sciences,
		D\'ubravsk\'a cesta 9, 842 28 Bratislava, Slovakia\\
	 $^{2}$ Department of Optics, Comenius University,
		Mlynsk\'a dolina, 842 15 Bratislava, Slovakia\\
         $^{3}$ Faculty of Informatics, Masaryk University,
                Botanick\'a 68a, 602 00 Brno, Czech republic
}
\maketitle

\begin{abstract}
Nature provides us with a restricted set of microscopic interactions.
The question is whether we can synthesize out of these fundamental
interactions an arbitrary unitary operator. In this paper we present
a constructive
algorithm for realization of any unitary operator
which acts on a (truncated) Hilbert space of a single bosonic mode.
In particular,  we consider a physical implementation
of unitary transformations acting on 1-dimensional vibrational states
of a trapped ion. As an example we present an algorithm which realizes
the discrete Fourier transform.
\end{abstract}
\pacs{03.65.Bz,42.50.Dv,32.80.Pj}
} 

\section{Introduction}

Controlled manipulations with individual quantum systems
such as trapped ions or cold atoms in atomic physics,  molecules
irradiated by laser fields, and Rydberg atoms interacting with
quantized micromaser fields, provides us with a deeper understanding
of fundamental principles of physics. Simultaneously, the possibility
to control individual quantum systems
opens new perspectives in application
of quantum physics.
Specifically, coherent control
over dynamics of quantum systems is of vital importance
for quantum computing and information
processing \cite{Steane,Vedral}.

Quantum information processing can be schematicically devided into
three stages. The first stage is the encoding of information into
quantum systems,
i.e. this corresponds to a preparation of states of quantum systems.
The second stage is the information processing which in general is
equivalent to a specific unitary evolution of the quantum system, i.e.
this is an application of a given quantum algorithm. The third stage
is the reading of output states of quantum registers (i.e. the ``decoding''
of information from quantum systems).
Obviously this final stage
is  the measurement of quantum system and the reconstruction of relevant
information.

There are several physical systems which are believed to be hot candidates
for quantum processors. In particular, Cirac and Zoller \cite{Cirac1995}
have shown that a system of trapped ions can be  utilized as a prototype
of a quantum computer.   Therefore it is of great interest to understand
how the three stages of the information processing as specified above
can be implemented in this system.

{\em (i) State preparation}: Recently,  several methods for
{\it deterministic} synthesis (preparation) of vibrational states of
trapped ions have been proposed. In particular,
a scheme for preparation of quantum states
of one and two-mode bosonic fields (e.g. 1 and 2-D quantum states of
vibrational motion of trapped ions) have been proposed
by Law and Eberly \cite{Law1996}, and Kneer and Law \cite{Kneer1998}
(see also \cite{Drobny1998} and for
a more general  discussion on the state preparation see \cite{special}).

{\em (ii) State measurement}:
There are various experimental techniques which allow to measure
and reconstruct quantum states of trapped ions (for a review see
\cite{Wineland}).

{\em (iii) Arbitrary unitary evolution}:
One of the most important task in information processing is to design
processors which take an {\em arbitrary} input and process it according
to a specific prescription.
Nature provides us with a restricted set of microscopic interactions.
The question is whether we can synthesize out of these fundamental
interactions an arbitrary unitary operator. 
In this paper we present a constructive
algorithm for realization of any unitary operator
which acts on a finite-dimensional Hilbert space.
An algorithmic proof that any discrete finite-dimensional unitary 
matrix can be factorized into a sequence of two-dimensional beam splitter
transformations was given by Reck et al.\cite{Reck1994}.
The problem of controlled dynamics
of quantum systems has been addressed recently by Harel and Akulin
\cite{Harel1999} and by Lloyd and Braunstein \cite{Lloyd1999}.
Harel and Akulin \cite{Harel1999}
have proposed a method to attain any desired unitary
evolution of quantum systems by switching on and off
alternatively two distinct constant perturbations.
The power of the method was shown in controlling
the 1-D translational motion of a cold atom.

Our aim is to find a constructive algorithm to realize an arbitrary unitary
operator $\hat{V}$ which transforms any state $|\psi\r$
of a single bosonic mode, e.g., an 1-D vibrational state of a trapped ion
in $x$ direction, to another state $|\psi'\r$, i.e.,
$|\psi' \r = \hat{V} |\psi\r$.
In particular, we consider a truncated ($N+1$)-dimensional Hilbert space
of the bosonic mode. Within this truncated Hilbert space the desired
unitary operator $\hat{V}$ in the number-state basis reads
\be 
\hat{V}=\sum_{m,n=0}^{N,N} V_{n,m}|n\r \l m|
\label{tr1}
.\ee 
Under action of the operator $\hat{V}$ a given input state
transforms as
\be 
|\psi\r=\sum_{m=0}^N c_m |m\r
{\buildrel \hat{V} \over \longrightarrow}
|\psi'\r=\hat{V}|\psi\r=\sum_{n=0} c'_n |n\r
\label{tr2}
\ee 
where
$c'_n=\sum_{m=0}^N V_{n,m} c_m$. Our task is to represent
by a feasible physical process any operator $\hat{V}$ (\ref{tr1}).
The synthesis of operators enables thus to realize universal quantum gates
for qubits which can be encoded into vibrational levels.

The paper is organized as follows. In Section \ref{sec.tools}
we briefly introduce physical tools which we use to realize arbitrary
unitary operators for a trapped ion.
The synthesis algorithm is described in Section \ref{sec.synthesis}.
The method is illustrated in Section \ref{sec.example} where a realization
of quantum gates which perform the Fourier transform is considered.
We also discuss stability of the algorithm and possibilities to realize
also non-unitary operators. We finish our paper with conclusions.

\section{Tools for synthesis: laser stimulated processes}
\label{sec.tools}

Our realization of the unitary transformation $\hat{V}$ given by
Eq.(\ref{tr1}) is based on an enlargement of the Hilbert subspace of
the given system.
Namely, the transformed bosonic mode corresponds to one vibrational
mode (in $x$ direction, for concreteness) of a quantized center--of--mass
motion of an ion confined in the 2-D trapping potential.
Within our synthesis procedure the vibrational $x$-mode becomes
entangled with the auxiliary degrees of freedom ({\it ancilla})
which are represented by the second vibrational mode
(e.g., quantized vibrational motion in $y$ direction)
and three internal electronic levels $|a\r, |b\r, |c\r$ of the ion.
The particular choice of the physical system is motivated by
a feasibility of highly coherent control over motional
degrees of freedom as demonstrated in recent experiments \cite{Wineland}
which makes trapped ions to be a hot candidates for quantum processors.

A physical realization of the desired operator $\hat{V}$ for an ion
confined in a 2-D trapping potential consists of a sequential switching
(on/off) of laser fields which irradiate the ion.
Namely, we utilize four types of laser stimulated interactions
which are associated with the following (effective)
interaction Hamiltonians:
\be 
\begin{array}{l}
\hat{H}^{(1,m)}=(\Delta_{y}+\hat{s}_{x}) (|a\r \l a| - |b\r \l b|)+
              g_1 |a\r \l b|+g_1^{\star} |b\r \l a|, \hspace{-1cm} \\
\hat{H}^{(2)}=
g_2 |b\r \l c| {\cal F}(\hat{a}_{y}^{\dagger}\hat{a}_{y})\hat{a}_{y}^{\dagger}+
g_2^{\star} \hat{a}_{y}{\cal F}(\hat{a}_{y}^{\dagger}\hat{a}_{y})|c\r \l b|,
\\
\hat{H}^{(3)}=
g_3 |b\r \l c|+g_3^{\star} |c\r \l b|, \\
\hat{H}^{(4)}=
g_4 |b\r \l c| {\cal F}(\hat{a}_{x}^{\dagger}\hat{a}_{x})\hat{a}_{x}^{\dagger}+
g_4^{\star} \hat{a}_{x} {\cal F}(\hat{a}_{x}^{\dagger}\hat{a}_{x})|c\r \l b|.
\end{array}
\label{ha4}
\ee 
where $\hat{s}_{x}=\sum_m s_m |m\r_x \l m|$ with
$s_{m}=\chi \left[ 1+ \mbox{e}^{-2\eta_x^2} L^0_{m}(4\eta_x^2) \right]$
and ${\cal F}(\hat{a}_q \hat{a}_q)=\mbox{e}^{-\eta_q^2/2}
\sum_{k=0} \frac{(-1)^k \eta_q^{2k}}{(k+1)!k!}
\hat{a}_q^{\dagger k} \hat{a}_q^k$.
The Lamb--Dicke parameters are defined as
$\eta_q=\omega_q/(c\sqrt{2m_a\nu_q})$ (assuming units such that $\hbar=1$)
where $\omega_q, \nu_q$ are frequencies of the laser and vibrational
mode in direction $q$ ($q=x,y$), respectively.
Further, $L^0_m$ denote the Laguerre polynomial;
$m_a$ is the mass of the ion.

The dynamical Stark shift operator $\hat{s}_{x}=\sum_m s_m |m\r_x \l m|$
is induced by a detuned standing--wave laser field applied
in $x$ direction \cite{Kneer1998}.
For $\chi\gg g_{1}$ the effective Hamiltonian $H^{(1,m)}$
addresses the states with fixed number $m$ of excitations
(indicated by the superscript) in the mode $x$ setting the detuning
$\Delta_y$ of the laser field, applied in the orthogonal $y$ direction,
equal to the dynamical Stark shift $-s_m$.
In other words, when the interaction is governed by the Hamiltonian
$H^{(1,m)}$ with $\Delta_{y}=-s_m$
there is an exchange of the population only between the states
$|m,n\r \otimes |a\r \Leftrightarrow |m,n\r \otimes |b\r$
with the given number of quanta $m$ in the vibrational mode $x$ and
any number of phonons $n$ in the mode $y$.
The populations of the other number states
(with number of excitations in the mode $x$ different from $m$)
effectively do not change due to a large detuning. However, there are
significant phase shifts of the amplitudes of the off-resonant states.
It should be stressed that the addressing of states with a given number $m$
of excitations in the mode $x$ is effective only for large ratios
$\chi/g_{1} \gg 1$. Moreover, the approximate Hamiltonian $H^{(1,m)}$
is itself justified only when $\chi\gg |g_{1}|$
(for details see \cite{Kneer1998}).

The considered interactions (\ref{ha4}) are quite typical for a trapped ion.
The effective interaction Hamiltonians represent 
a classical driving of the ion [see $H^{(3)}$] and 
a non-linear Jaynes-Cummings model \cite{Vogel1995}
[see $H^{(2),(4)}$] when the lasers 
applied in given directions ($x$,$y$) are tuned to appropriate vibrational 
sidebands. These Hamiltonians are thoroughly discussed in
papers \cite{Kneer1998,Gardiner1997}.
   
The dynamics governed by the interaction Hamiltonians
(\ref{ha4}) can be separated into independent 2-D subspaces.
Switching on a particular interaction ``channel''
associated with one of the interaction Hamiltonians $\hat{H}^{(p)}]$
(\ref{ha4}) for a time $\tau$ is described as the action of the corresponding
unitary time-evolution operator
$\hat{U}^{(p)}=\exp[-i \hat{H}^{(p)} \tau]$ on the state vector
of the system under consideration.

\section{Synthesis of transformations}
\label{sec.synthesis}

To implement the desired transformation $\hat{V}$ (\ref{tr1})
for an ion confined in the 2-D trapping potential
we realize a mapping
$|\psi_x,0_y\r {\buildrel \hat{V} \over \longrightarrow}
|0_x,\psi'_y\r=|0_x, \hat{V}\psi_y\r$
of two--mode bosonic states
(here subscripts indicate particular vibrational modes;
in what follows the subscripts will be omitted for given ordering of modes).
To be more explicit, the realization of the transformation $\hat{V}$
can be expressed as the mapping in the extended Hilbert space
of two bosonic modes and internal electronic levels in the following form:
\be 
|0,\psi'\r \otimes |b\r = {\cal N} \hat{P}_{|b\r}\,
\hat{B}\, \hat{A}\, |\psi,0 \r \otimes |a\r
\label{st1}
\ee 
where the operators $\hat{A}$ and $\hat{B}$ represent
a sequence of four types of unitary operations.
The final projection  $\hat{P}_{|b\r}$ on the state $|b\r$
selects {\em conditionally} the right outcome (${\cal N}$ is a proper
normalization constant).
In the subsequent step one could use the two--mode linear coupler
based on laser stimulated Raman transitions \cite{Steinbach1997}
to swap the states of the vibrational modes,
i.e., $|0,\psi' \r \rightarrow |\psi',0 \r$.
The additional $\pi$-pulse can be used to flip the electronic state
from the level $|b\r$ into the initial level $|a\r$.

The operators $\hat{A}$, $\hat{B}$, $P_{|b\r}$ appearing
in the Eq.(\ref{st1}) indicate four essential steps which lead us
subsequently to the desired transformations.

\begin{figure}[t]
\centerline{\epsfig{width=8.0cm,file=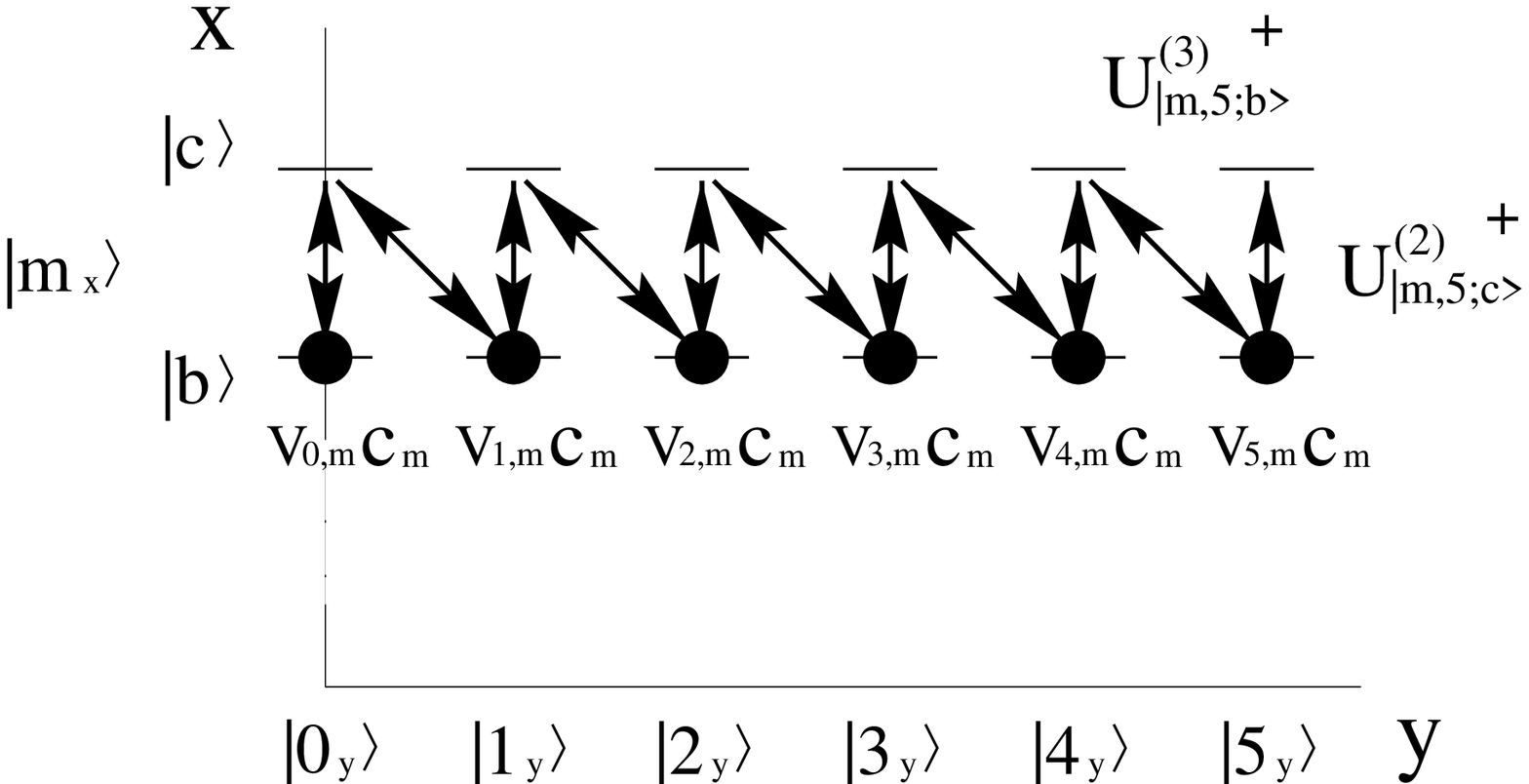}}
\caption{The action of the operator $\hat{A}^\dagger(m)$
on the row with a fixed number $m$ of phonons in the mode $x$.
The population from the superposition
$\sum_n V_{n,m} |n\r_y$ is ``swept'' down into the vacuum $|0\r_y$ in the
mode $y$.
}
\label{fig1}
\end{figure}

\noindent {\bf Step A}.
The operator $\hat{A}$ ``spreads'' the amplitudes ${c}_{m}$'s of
the component number states $|n_{x},0\r\otimes |a\r$ over
the whole Hilbert space
${\cal H}_{x}\otimes {\cal H}_{y}\otimes {\cal H}_{in}$
so that the entangled state of the composed system becomes\footnote{
In our synthesis algorithm we neglect off-resonant transitions            
between internal levels in $\hat{H}^{(1,m)}$. Strictly speaking,          
the equality sign applies only in the limit $\chi/|g_1| \to\infty$.}
\be 
|\psi^A \r = \sum_{m,n=0}^{N,N} V_{n,m} {c}_{m} \mbox{e}^{i\varphi_m^A}
|m,n\r \otimes |a\r
.\label{sa1}
\ee 
This task can be done by the method of 1-D quantum state synthesis
proposed by Law and Eberly \cite{Law1996}.
An important tool represents also the photon--number dependent
interaction $\hat{H}^{(1,m)}$ considered by Kneer and Law \cite{Kneer1998}
which enables to address the subspaces with a fixed number $m$ of phonons
in the $x$ direction.
The operator $\hat{A}$ can be written as
\be 
\hat{A}=\prod_{m=0}^{N}
\hat{U}^{(1,m)}_{-\pi/2\exp(i\phi_m)}\hat{A}(m)\hat{U}^{(1,m)}_{\pi/2}
\ee 
where
\be 
\hat{A}(m)=\hat{U}^{(2)}_{|m,N;b\r } \hat{U}^{(3)}_{|m,N-1;c\r } \ldots
        \hat{U}^{(2)}_{|m,1;b\r }\hat{U}^{(3)}_{|m,0;c\r }
.\label{aa1}
\ee 
The subscript of the unitary transformation $\hat{U}^{(1,m)}_{g_1 \tau}$
indicates the required setting of the corresponding interaction parameter
$g_1 \tau=|g_1 \tau |\mbox{e}^{i\phi_m}$.
In other transformations $\hat{U}^{(p)}_{|\Phi\r}$ ($p=2,3$)
the subscripts denote steps in 1D quantum-state synthesis as explained below.
In other words,
within a particular subspace with a fixed number $m$ of phonons in the $x$
direction we flip from the electronic level $|a\r$ to $|b\r$
by means of $\hat{U}^{(1,m)}_{\pi/2}$.
Then to ``spread'' $|m,0\r \otimes |b\r$ we apply 1-D quantum state
synthesis associated with the action of the operator $\hat{A}(m)$
to ``prepare'' the superposition
$\sum_{n=0}^{N} V_{n,m} |m,n\r \otimes |b\r$.
After that the electronic level $|b\r$ is flipped back to $|a\r$
via the action of $\hat{U}^{(1,m)}_{-\pi/2\exp(i\phi_m)}$
[here $\phi_m$'s represent proper phase factors which will be discussed
later, see Eq.(\ref{phim})].

The appropriate interaction parameters for our 1-D quantum state synthesis
can be found when we solve the inverse task which is given by the inverse
transformation:
$|m,0\r \otimes |b\r = \hat{A}^{\dagger}(m) \sum_n V_{n,m} |m,n\r \otimes |b\r$.
The inverse task is based on "sweeping" down the probability
from the component states of the given superposition in the $y$ mode
into the vacuum. Therefore the subscripts
of the unitary operators $\hat{U}^{(p)}_{|\Phi\r}$ ($p=2,3$) in (\ref{aa1})
indicate that interaction parameters $g_p \tau=e^{i\phi}|g_p \tau|$
have to be chosen in such way that after the action of
the $\hat{U}^{\dagger (p)}_{|\Phi\r}$
the amplitude (population) of the component state $|\Phi\r$
becomes equal to zero.
The action of the operator $\hat{A}^\dagger(m)$ is shown
schematically in Fig.~\ref{fig1}.
We have applied the procedure introduced by Law and Eberly
for synthesis of 1-D bosonic states in the straightforward way.
Therefore we refer readers  to the original paper \cite{Law1996}
for other details (generalized Hamiltonians to operate beyond
the Lamb--Dicke limit can be found in \cite{Gardiner1997}).
Note that in our case we apply the state--synthesis procedure only in
$y$ direction and the resulting state (\ref{sa1})
remains {\em unknown}.

To resume, the action of the operator $\hat{A}$ encodes
the matrix elements $V_{n,m}$
(multiplied with the amplitudes $c_m$'s of unknown state)
into rows of the 2D vibrational ``lattice'' of number states $|m,n\r$.
In the next step an appropriate superposing of columns
within the 2D vibrational number-state ``lattice'' is required.

\begin{figure}
\centerline{\epsfig{width=8.cm,file=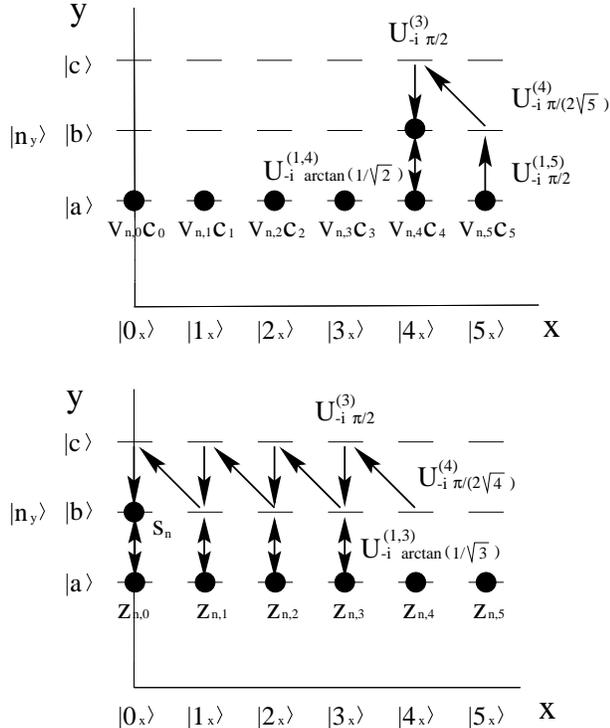}}
\caption{
The action of the operator $\hat{B}$ on the column with $n$
vibrational quanta in the mode $y$.
The same changes occur simultaneously on ``parallel'' columns
with different $n$. The upper figure shows
the action of $\hat{U}^{(1,N)}_{-i \pi/2}$ (for $N=5$) followed
by the basic sequence of operations $\hat{B}(m)$ (for $m=4$).
Decreasing the number of quanta $m$ in $x$ direction
the basic sequence $\hat{B}(m)$ is recursively repeated
as shown in the lower figure.
After the action of the operator $\hat{B}$ the value of the amplitude
$s_n=(\sum_{m=0}^{N} V_{n,m} c_m)/\protect\sqrt{N+1}$.
}
\label{fig2}
\end{figure}

\medskip
\noindent {\bf Step B.}
In the second step the operator $\hat{B}$ creates the state in which the
amplitudes of the states $|0,n\r \otimes |b\r$ $(n=0,...N)$ are proportional
to $c'_n=\sum_{m=0}^{N} V_{n,m}\ c_{m}$.
The state of the system $|\psi^B\r = \hat{B} |\psi^A\r$
after this synthesis step reads
\be
|\psi^B\r =
\frac{1}{\sqrt {N+1}} |0,\psi'\r \otimes |b\r +
   \sum_{m=0}^{N}\sum_{n=0}^{N} z_{m,n}|m,n\r \otimes |a\r.
\hspace{-1.0cm} \nonumber \\
\label{st3}
\ee
The operator $\hat{B}$ can be written in the form
\be
\hat{B}=\left[ \prod_{m=0}^{N-1} \hat{B}(m) \right] \hat{U}^{(1,N)}_{-i \pi/2}
\label{bb1}
\ee
where
\be
\hat{B}(m)=\hat{U}^{(1,m)}_{-i \arctan(1/\sqrt{N+1-m})  }
    \hat{U}^{(3)}_{-i \pi/2}  \hat{U}^{(4)}_{-i \pi/(2 \sqrt{m+1})}
.\label{bb2}
\ee
Here the subscripts of the unitary operators $U_{g_p \tau}^{(p)}$
indicate again the proper choice of interaction parameters
$g_p \tau$.\footnote{
Here and in Fig.~\ref{fig2} the required setting $g_4 \tau=
-i \pi/(2 \sqrt{m+1})$ refers for clarity to the Lamb-Dicke regime 
$\eta_x\ll 1$. Outside of the Lamb-Dicke regime the Rabi frequency
$\sqrt{m+1}$ is simply replaced by the nonlinear Rabi frequency 
$\mbox{e}^{-\eta_x^2/2} L_{m}^{1}(\eta_x^2)/\sqrt{m+1}$ 
where $L^{1}_{m}$ denotes the associated Laguerre polynomial.
}

The action of the operator $\hat{B}$ on a particular column with a given
number $n$ of phonons in the $y$ mode is shown schematically
in Fig.~\ref{fig2}.
Simultaneously, the operator $\hat{B}$ acts in the same way on ``parallel''
columns with different $n$.
As illustrated in the upper part of Fig.~\ref{fig2},
the operator $\hat{U}^{(1,N)}_{-i \pi/2}$ (for $N=5$)
transfers the population from the internal level $|a\r$ to $|b\r$
only in the row with the fixed number of quanta $m=5$ in the mode $x$.
Further, this population is transfered to the internal level $|b\r$
in the neighboring row with the number of quanta $m=4$ performing
transformations
$\hat{U}^{(3)}_{-i \pi/2} \hat{U}^{(4)}_{-i \pi/(2 \sqrt{m+1})}$.

The equal superposition of the amplitudes
$V_{n,4} c_4$ and $V_{n,5} c_5$ in the row with $m=4$
is obtained after action of $\hat{U}^{(1,4)}_{-i \arctan(1/\sqrt{2})}$,
i.e. undergoing one half of the Rabi flipping.
Decreasing $m$ (the number of quanta in the mode $x$)
the basic sequence $\hat{B}(m)$ is recursively repeated
as indicated in the lower part of Fig.~\ref{fig2}.
Note that $\hat{U}^{(1,m)}_{-i \arctan(1/\sqrt{N-m+1})}$,
is responsible for adding of the amplitude $V_{n,m} c_m$
(with a proper weight) to previously superposed amplitudes
$\sum_{k=m+1}^N V_{n,k} c_k$ performing an adequate part of the Rabi flipping.
After the action of the whole operator $\hat{B}$ the value of the amplitude
$s_n=\frac{1}{\sqrt{N+1}}\sum_{m=0}^{N} V_{n,m} c_m$.
The transformed state $|\psi^B\r$ is given in Eq.(\ref{st3}).

At this step we should notice that {\it each} action of the ``elementary''
unitary operator $\hat{U}^{(1,m)}$, associated with
the interaction Hamiltonian $\hat{H}^{(1,m)}$ with $\Delta_y=-s_m$,
causes significant phase shifts on off-resonant rows
with the number of quanta in the mode $x$ different from $m$ \cite{Kneer1998}.
These phase shifts have to be compensated {\it in advance} in order
to ``superpose'' the amplitudes via the action of $\hat{B}(m)$
as described above
[see the r\^{o}le of $\hat{U}^{(1,m)}_{-i \arctan(1/\sqrt{N-m+1})}$].
This compensation can be done when we include appropriate phase shifts
$\phi_m$'s directly in the operator $\hat{A}$ [see Eq.({\ref{aa1})].
The explicit expression reads
\be 
\phi_m=-\frac{\pi}{2}-\!\!\sum_{k=0}^{m-1} f_m^{(k)}(\pi)
       -\!\!\!\!\sum_{k=m+1}^{N} f_m^{(k)}(\arctan\frac{1}{\sqrt{N-k+1}})
\hspace{-1.0cm} \nonumber \\
\label{phim}
\ee 
where
$f_m^{(k)}(|g_1| \tau)=\arg[\cos(\Omega_m^{(k)} \tau)+
      i \frac{s_k-s_m}{2\Omega_m^{(k)}} \sin(\Omega_m^{(k)} \tau)]$
with 
$\Omega_m^{(k)}=\sqrt{(s_k-s_m)^2/4+|g_1|^2}$.
The origin of the expression (\ref{phim}) can be traced back
to the operators $\hat{U}^{(1,m)}$ in the steps $\hat{A}$ and $\hat{B}$.
The aim is to cancel the phase shifts of the amplitudes
in order to ``superpose'' them on the $m$-th row
by means of $\hat{B}(m)$.
Therefore the first sum in (\ref{phim}) compensates (in advance)
the subsequent shifts in $\hat{A}$ due to $\hat{U}^{(1,k)}$
for $k=0,\ldots,m-1$. The second sum compensates the shifts which will
take place during ``superposing'' operations
$\hat{B}(k)$ for $k=m+1,\ldots,N$ which precede $\hat{B}(m)$.

\medskip
\noindent {\bf Step C.}
Comparing the state $|\psi^B\r$ [Eq.(\ref{st3})] with the desired one
[Eq.(\ref{st1})] we see that the target state is entangled to the internal
level $|b\r$. On the other hand,
also undesired component states $|m\neq 0,n\r \otimes |a\r$
are now contributing into Eq.(\ref{st3})
with nonvanishing amplitudes.
However (fortunately), all the undesired components are entangled
with the internal level $|a\r$.
Therefore we can perform a conditional measurement to project
the state vector (\ref{st3}) on the internal level $|b\r$.
To be more specific, the internal state of the ion can be determined
by driving transition from the level
$|a\r$ to an auxiliary level $|r\r$
and observing the fluorescence signal\footnote{
As a check one could drive also transition from the level $|c\r$       
to another auxiliary level $|r'\r$. Errors in the synthesis procedure  
are thus indicated by presence of the fluorescence signal.}. 
No signal (no interaction with probing field) means
that the undisturbed ion is occupying the level $|b\r$
being in the motional state $|0,\psi'\r $.
It means that after the conditional measurement
[indicated in (\ref{st1}) by the projector $\hat{P}_{|b\r}$]
the state vector (\ref{st3}) is reduced to the desired state vector
(\ref{st1}).
The probability to find the right outcome for the unitary transformations
is equal to $\frac{1}{N+1}$.

In spite of the involved {\em conditional selection} of the right outcomes,
our algorithm is {\em universal} as the sequence of
the ``elementary'' operations (with appropriate interaction parameters)
which represents the desired transformation is always {\em independent}
of input states.
Moreover, in contrary to conditional measurement schemes
known from quantum state preparation, in our case
the probability of the right outcome is constant
being also {\em independent} of input states.

\section{Discussion}
\label{sec.example}

One of important applications of the operator synthesis
is a realization of  {\em universal} quantum gates for qubits
which are encoded in vibrational levels.
The number states of the vibrational mode can represent
a quantum register.

To illustrate our synthesis procedure we considered a realization
of the operator which ``rotates'' the population between
$N$ vibrational Fock states under consideration.
\be
\hat{V}_R=\frac{1}{\sqrt{N+1}}
( \sum_{j=0}^{N-1} |j+1\r \l j| + |0\r \l N| )
.\ee
It corresponds to a cyclic ``rotation'' of the quantum register.
The operator $\hat{V}_R$ represents the unitary exponential phase
operator of the Pegg--Barnett formalism \cite{Pegg1989}.

In the second example the synthesis procedure is applied for
the unitary operator of the quantum Fourier transform
defined as \cite{Barenco1996}
\be
\hat{V}_{QFT}=\frac{1}{\sqrt{N+1}}
\sum_{m,n=0}^{N-1} \exp\left(2\pi i \frac{m n}{N+1}\right) |m\r \l n|
.\ee
The operator of the quantum Fourier transform represents an important tool
in quantum computing \cite{Barenco1996}.

\begin{figure}
\centerline{\epsfig{width=8.0cm,file=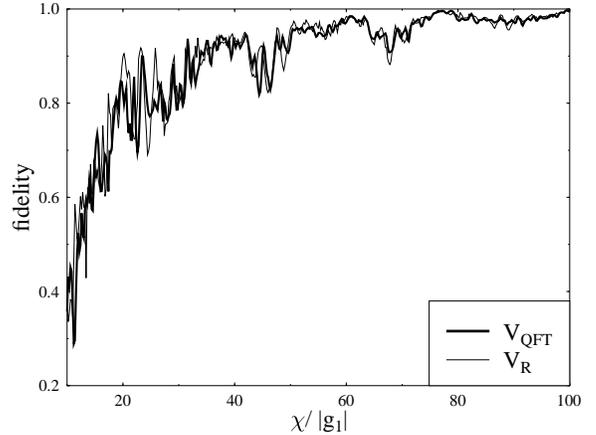}}
\caption{
The fidelity of outputs to the ideally transformed states
as function of the ratio $\chi/|g_1|$ for
the quantum Fourier transform $\hat{V}_{QFT}$ (thick solid line)
and the exponential phase operator $\hat{V}_{R}$ (thin solid line).
Lamb--Dicke parameters are $\eta_x=\eta_y=0.4$.
}
\label{fig3}
\end{figure}

In the presented synthesis procedure we have neglected
transitions between internal levels on off-resonant ``rows''
when the interaction Hamiltonian $\hat{H}^{(1,m)}$ is applied.
Strictly speaking, the off-resonant transitions in $\hat{H}^{(1,m)}$
can be neglected only in the limit $\chi/|g_1|\to \infty$.
Our estimation of the error due to the finite values of the ratio
$\chi/|g_1|$ (feasible in practice) is based on the fidelity of the
outputs to the ideally transformed states (\ref{st1}).
The fidelity of two states $|\Phi\r$,
$|\Phi^\prime \r$ is defined as their squared scalar product
$\vert \l \Phi^\prime | \Phi\r \vert^2$.
As a testing input state of the $x$-mode we can take a uniform
superposition of involved number states, i.e.,
$|\psi\r = \frac{1}{\sqrt{N+1}} \sum_{m=0}^N |m\r_x$
(the initial internal state of the ion is $|a\r$ and the vibrational
$y$-mode is in the vacuum).
The figure~\ref{fig3} shows that the fidelity is close to one for large
ratios $\chi/|g_1|\gg 1$ for both operators $\hat{V}_R$ and 
$\hat{V}_{QFT}$.\footnote{
In Fig.~\ref{fig3} we consider parameters outside of the Lamb--Dicke
regime. To operate in the Lamb--Dicke regime requires 
a further increase of the ratio  $\chi/|g_1|\gg 1$ to reach a fidelity close
to one.}        
In such case both the validity of the approximate
Hamiltonian $\hat{H}^{(1,m)}$ is justified \cite{Kneer1998} and
the level flipping ($|a\r \leftrightarrow |b\r$) on off-resonant
``rows'' (as the source of non-ideal fidelity) can be neglected.

Let us note that for nonunitary transformations the probability
depends on initial states and can be from the interval $(0,1)$.
However, the realization of the transformation
(given by the sequence of the ``elementary'' operations
with appropriate interaction parameters) is always {\em independent}
of input states.

\section{Conclusions}

In this paper we have proposed a constructive algorithm for
synthesis of operators, a superior task to synthesis of quantum states.
Our method allows us to find
{\em analytical} expressions for switching times
and interaction parameters of the utilized laser stimulated processes
via which an arbitrary unitary dynamics can be realized.
One of important applications of the operator synthesis
is a realization of  {\em universal} quantum gates for qubits
which are encoded in vibrational levels. As an example we consider
realization of the discrete Fourier transform.

The solution of the problem we present in our paper is neither unique nor
optimal (in a sense of number of elementary operations used for a
construction of the given unitary operator). The optimization of the
procedure is the problem which has to be solved. The other problem which
deserves attention is the stability of the algorithm with respect to noise
inherent in the system. In fact, one can consider two types of uncertainties
which might play an important r\^ole. Firstly, it is the noise induced by
the environment, i.e. the elementary gates are not unitary. The second
source of noise (kind of technical noise) is due to the fact that it is
not possible to keep the interaction times and parameters to be fixed
as given by the theory. Fluctuations in these parameters might reduce
the fidelity of the realization of the desired unitary evolution.
We will address these questions elsewhere.

\acknowledgements
We thank Gil Harel and Vladimir Akulin for correspondence, and
Jason Twamley for discussions.
This work was supported in part by the Slovak Academy of Sciences (project
VEGA), by the GACR (201/98/0369), and by the Royal Society.


\end{document}